\documentclass[fleqn,12pt,twoside]{article}
\usepackage{espcrc1}
\usepackage{t1enc,wrapfig}

\usepackage{graphicx}
\usepackage[figuresright]{rotating}

\title{Molecular resonance phenomena and alpha-clustering: recent
progress and perspectives} 

\author{C. Beck\address[IReS]{Institut de Recherches Subatomiques,
UMR7500, IN2P3-CNRS/Universit\'e Louis Pasteur, B.P. 28, F-67037
Strasbourg Cedex 2, France} }

\begin{document}

\maketitle

\begin{abstract}

{\small The connection between molecular resonance phenomena in light
heavy-ion collisions, alpha-clustering and extremely deformed states
in light $\alpha$-like nuclei is discussed. For example, the
superdeformed bands recently discovered in light N=Z nuclei such as
$^{36}$Ar, $^{40}$Ca, $^{48}$Cr, and $^{56}$Ni by $\gamma$-ray
spectroscopy may have a special link with resonant states in
collisions with $\alpha$-like nuclei. The resonant reactions
involving identical bosons such as $^{12}$C+$^{12}$C,
$^{16}$O+$^{16}$O $^{24}$Mg+$^{24}$Mg and $^{28}$Si+$^{28}$Si are of
interest. For instance, a butterfly mode of vibration of the
J$^{\pi}$ = 38$^{+}$ resonance of $^{28}$Si+$^{28}$Si has been
discovered in recent particle $\gamma$-ray angular correlations
measurements. The search for signatures of strongly deformed shapes
and clustering in light N=Z nuclei is also the domain of charged
particle spectroscopy. The investigation of $\gamma$-decays in
$^{24}$Mg has been undertaken for excitation energies where
previously nuclear molecular resonances were found in
$^{12}$C+$^{12}$C collisions. In this case the $^{12}$C-$^{12}$C
scattering states can be related to the breakup resonance and,
tentatively, to the resonant radiative capture $^{12}$C+$^{12}$C
reaction. } 

\end{abstract}

\section{Introduction}

The recent discovery of highly deformed shapes and superdeformed (SD)
rotational bands in the N=Z nuclei $^{36}$Ar~\cite{Svensson00},
$^{40}$Ca~\cite{Ideguchi01}, $^{48}$Cr~\cite{Lenzi96,Thummerer01} and
$^{56}$Ni~\cite{Rudolph99} has renewed the interest in theoretical
calculations for {\it sd}-shell nuclei around
$^{40}$Ca~\cite{Inakura02,Sakuda02,Kanada02}. Therefore the A$_{CN}$
$\approx$ 30-60 mass region becomes of particular
interest~\cite{Beck04} since quasimolecular resonances have also
been observed for these $\alpha$-like nuclei, in particular, in the
$^{28}$Si+$^{28}$Si reaction~\cite{Betts81,Beck01,Nouicer99}.
Although there is no experimental evidence to link the SD bands with
the higher lying rotational bands formed by known quasimolecular
resonances~\cite{Bromley60}, both phenomena are believed to originate
from highly deformed configurations of these systems. The
interpretation of resonant structures observed in the excitation
functions in various combinations of light $\alpha$-cluster nuclei in
the energy regime from the barrier up to regions with excitation
energies of 30-50~MeV remains a subject of contemporary debate. In
particular, in collisions between two $^{12}{\rm C}$ nuclei, these
resonances have been interpreted in terms of nuclear
molecules~\cite{Bromley60}. However, in many cases these structures
have been connected to strongly deformed shapes and to the
alpha-clustering phenomena~\cite{Marsh86,Flocard84,Leander75}. There
has been a continuing discussion as to whether these molecular
resonances represent true cluster states in the $^{24}$Mg compound
system, or whether they simply reflect scattering states in the
ion-ion potential. In this paper, a few examples will be given
showing the close connection between molecular resonance phenomena,
alpha-clustering, and nuclear deformation. 

\section{Molecular resonances in $^{28}$Si+$^{28}$Si}

	\begin{wrapfigure}[32]{r}[0cm]{8cm}
	\includegraphics[width=8cm,height=12cm]{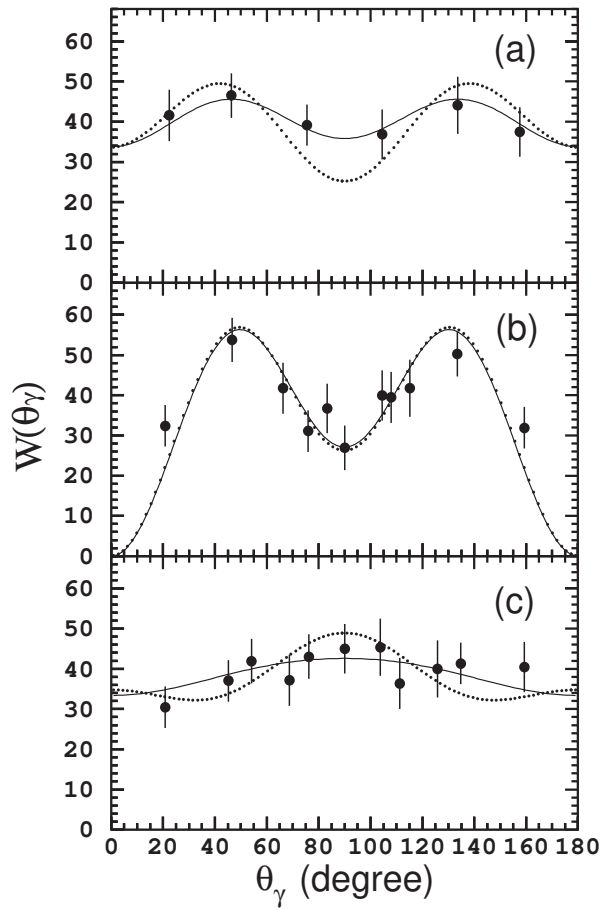}
	\parbox{80mm}
		{    \caption{\small\em $\gamma$-ray angular 
                                         correlations of the   
		 (2$_{1}^{+}$,2$_{1}^{+}$) state of the 
                 $^{28}$Si + $^{28}$Si
		 exit-channel for the 3 quantization axes defined 
                 in the text. The
		 solid and dashed curves are fits of the data and 
                 predictions, respectively.}
		{\label{}   
		{}}}
	\end{wrapfigure}

The molecule-like sequences of resonances observed in $^{28}$Si +
$^{28}$Si with measured angular momenta up to L = 42$\hbar$
represented some nuclear excitations with the highest spins ever
observed~\cite{Betts81}. In the number of open channels
model~\cite{Beck94,Beck95}, highly successful in selecting the
systems showing resonance behavior, the main condition for observing
a resonance behavior is associated with surface transparency. The
$^{28}$Si + $^{28}$Si reaction is a particularly favorable
case~\cite{Beck95}, where the corresponding optical model (OM)
potentials have small imaginary components at distances corresponding
to peripheral collisions. Therefore, the well established J$^{\pi}$ =
38$^{+}$ molecular resonance observed in $^{28}$Si + $^{28}$Si data
at E$_{lab}$ = 112~MeV has been studied at the {\sc Vivitron} Tandem
facility of the IReS by both fragment-fragment and
fragment-fragment-$\gamma$ coincidence 
measurements~\cite{Beck01,Nouicer99}. A subsequent
experiment~\cite{Chandana02} using charged particle spectroscopy
techniques with the {\sc Icare} multidetector array~\cite{Rousseau02}
has indicated the occurence of strongly deformed shapes at high spin
(with axis ratios consistent with SD bands) for the $^{56}$Ni
composite system at the resonant energy in agreement with very recent
$\gamma$-ray spectroscopy data obtained at much lower
spins~\cite{Rudolph99}. From the analysis of the particle angular
distributions of the mass-symmetric $^{28}$Si + $^{28}$Si
exit-channel~\cite{Nouicer99}, it could be concluded that, at the
resonance energy, the spin vectors of the $^{28}$Si fragments do not
couple with the orbital angular momentum, leading to $m=0$. The
fragment-fragment-$\gamma$ coincidence data~\cite{Beck01} demonstrate
that for the $^{28}$Si fragments, the mutually excited states are the
most strongly populated. The resonance behavior appears to involve
preferentially the low-lying states of the mass-symmetric channel. 

In Fig.~1 the results of the $\gamma$-ray angular correlations for
the mutual excitation exit-channel (2$_{1}^{+}$,2$_{1}^{+}$) are
shown. The distributions are presented in terms of the polar angles
with respect to the three different quantization axes defined as:~(a)
the beam axis, (b) the axis normal to the scattering plane, and (c)
the axis perpendicular to both axes defined in (a) and (b). The (c)
axis corresponds approximately to the molecular axis of the outgoing
binary fragments. the strong minimum in Fig.~1(b) at 90$^{\circ}$
implies that the magnetic substate m is equal to zero ($m=0$), and,
thus, that the intrinsic spin vectors of the 2$^{+}$ states are
oriented in the reaction plane perpendicularly to the orbital angular
momentum. The value of the total angular momentum, therefore, remains
close to ${\bf L} = 38\,\hbar$, in good agreement with the angular
distributions results~\cite{Nouicer99}. These observations do favor
the calculations of the molecular model of Uegaki and
Abe~\cite{Uegaki94}, which predict a vanishing spin alignment for
$^{28}$Si + $^{28}$Si arising from a butterfly mode of vibration of
the di-nuclear system at the resonance energy. 

\section{Spin-alignment measurements of molecular states}

The present $^{28}$Si + $^{28}$Si data show, for the first time in a
heavy-ion collision, a vanishing spin alignment. Therefore, the
comparison between the three symmetric systems $^{12}$C + $^{12}$C,
$^{24}$Mg + $^{24}$Mg and $^{28}$Si + $^{28}$Si in Fig.~2 shows an
interesting contrast in the spin orientation at resonance energies.

        \begin{wrapfigure}[30]{r}[0cm]{8cm}
        \includegraphics[width=8cm,height=12cm]{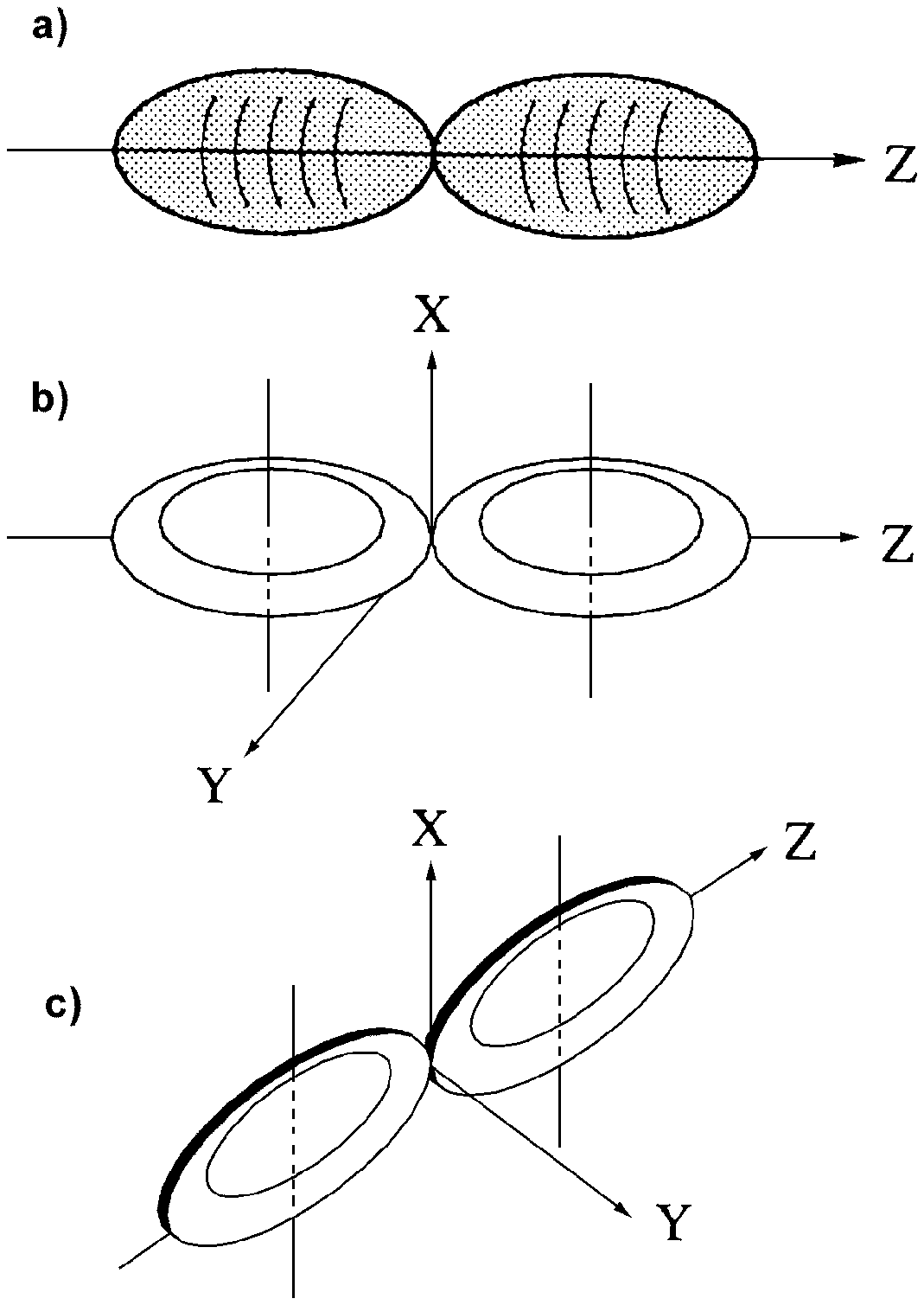}
        \parbox{80mm}
                {    \caption{\small\em
Equilibrium configurations of three different dinuclear systems : a)
for $^{24}$Mg + $^{24}$Mg, b) for $^{28}$Si + $^{28}$Si, and c) for
$^{12}$C + $^{12}$C.}
                {\label{}
                {}}}
        \end{wrapfigure}

The results indicate that the $^{28}$Si + $^{28}$Si oblate-oblate
system, illustrated by Fig.~2(b), is characterised by spin
disalignment in contrast to the spin alignment observed for both the
$^{12}$C + $^{12}$C system~\cite{Konnerth85,Wuosmaa03} of Fig.~2(c)
(oblate-oblate) and the $^{24}$Mg + $^{24}$Mg system~\cite{Wuosmaa87}
of Fig.~2(a) (prolate-prolate). Molecular-model
calculations~\cite{Uegaki94} are capable to explain 
the vanishing spin alignment in the oblate-oblate $^{28}$Si +
$^{28}$Si system~\cite{Nouicer99}, where both nuclear spin vectors
are perpendicular to the orbital angular momentum lying in the
reaction plane. 

A new butterfly mode of vibration for the $^{28}$Si-$^{28}$Si
triaxial molecule can then be speculated. However, the question of
why the spin orientations in the two oblate-oblate systems $^{28}$Si
+ $^{28}$Si and $^{12}$C + $^{12}$C are so different is still
unclear. Differences in the interactions between the constituent
nuclei may play a key role. For example, there is a remarkable
difference in the available molecular configurations in the
excitation spectra of the $^{12}$C and $^{28}$Si nuclei. In $^{12}$C
+ $^{12}$C there are few molecular configurations, located at the
energies associated with the observed resonances, while in $^{28}$Si
+ $^{28}$Si there are many more configurations. It is possible that
the large number of available configurations~\cite{Beck95} allows for
the formation of coherent or collective states. This explains the
sharp resonances which decay into many inelastic channels. In the
$^{12}$C + $^{12}$C system, such coherent effects may not be allowed
to develop. In this case the individual configurations may be
observed. Overall, the results of these experiments provide further
support for recent theoretical investigations that view the
resonances in terms of shape-isomeric states stabilized in
hyperdeformed secondary minima~\cite{Marsh86,Flocard84,Leander75}.
However, a more global understanding will surely require a more
fundamental synthesis of the theories which describe reaction
mechanisms with those describing nuclear structure at high excitation
energy and angular momentum. 

\section{$^{24}$Mg breakup states and the $^{12}$C+$^{12}$C molecule}

In the framework of the search for nuclear molecules the most
spectacular results have often been obtained for the
$^{12}$C+$^{12}$C reaction~\cite{Beck94}. However, the question
whether $^{12}$C+$^{12}$C molecular resonances represent true cluster
states in the $^{24}$Mg compound system, or whether they simply
reflect scattering states in the ion-ion potential is still
unresolved~\cite{Bromley60}. In many cases these structures have been
connected to strongly deformed shapes and to the alpha-clustering
phenomena, predicted from the $\alpha$-cluster model~\cite{Marsh86},
Hartree-Fock calculations~\cite{Flocard84}, the Nilsson-Strutinsky
approach~\cite{Leander75}. 
Various decay branches from the highly
excited $^{24}$Mg$^*$ nucleus, including the emission of $\alpha$
particles or heavier fragments such as $^{8}$Be and $^{12}$C, are
possibly available. However, $\gamma$-decays have not been observed
so far. Actually the $\gamma$-ray branches are predicted to be rather
small at these excitation energies, although some experiments have
been reported~\cite{McGrath81,Metag82,Haas97}, which have searched
for these very small branches expected in the range of
10$^{-4}$-10$^{-5}$ fractions of the total
width~\cite{Beck04,Beck03}. The rotational bands built on the
knowledge of the measured spins and excitation energies can be
extended to rather small angular momenta, where finally the
$\gamma$-decay becomes a larger part of the total width. The
population of such states in $\alpha$-cluster nuclei, which are lying
below the threshold for fission decays and for other particle decays,
is favored in binary reactions, where at a fixed incident energy the
composite nucleus is formed with an excitation energy range governed
by the two-body reaction kinematics. These states may be coupled to
intrinsic states of $^{24}$Mg$^{*}$ as populated by a breakup process
(via resonances) as shown in previous 
works~\cite{Fulton86,Curtis95,Singer00}. The $^{24}$Mg+$^{12}$C
reaction has been extensively investigated by several measurements of
the $^{12}$C($^{24}$Mg,$^{12}$C$^{12}$C)$^{12}$C breakup
channel~\cite{Fulton86,Curtis95,Singer00}. Sequential breakups are
found to occur from specific states in $^{24}$Mg at excitation
energies ranging from 20 to 35 MeV, which are linked to the ground
state and also have an appreciable overlap with the $^{12}$C+$^{12}$C
quasi-molecular configuration. Several attempts \cite{Curtis95} were
made to link the $^{12}$C+$^{12}$C barrier resonances
\cite{Bromley60} with the breakup states. The underlying reaction
mechanism is now fairly well established~\cite{Singer00} and many of
the barrier resonances appear to be correlated indicating that a
common structure may exist in both instances. This is another
indication of the possible link between barrier resonances and
secondary minima in the compound nucleus. The study of
particle-$\gamma$ coincidences in binary reactions in reverse
kinematics is probably a unique tool for the search for extreme
shapes related to clustering. In this way the $^{24}$Mg+$^{12}$C
reaction has been investigated with high selectivity at
E$_{lab}$($^{24}$Mg) = 130 MeV with the Binary Reaction Spectrometer
(BRS) in coincidence with {\sc Euroball IV} installed at the {\sc
Vivitron}~\cite{Beck04,Beck03}. The choice of the $^{12}{\rm C}(^{24}{\rm
Mg},^{12}{\rm C})^{24}{\rm Mg^{*}}$ reaction implies that for an
incident energy of E$_{lab}$ = 130~MeV an excitation energy range up
to E$^{*}$ = 30~MeV in $^{24}$Mg is covered~\cite{Curtis95}. The BRS
gives access to a novel approach to the study of nuclei at large
deformations~\cite{Beck04,Beck03}. The excellent channel selection
capability of binary and/or ternary fragments gives a powerful
identification among the reaction channels, implying that {\sc
Euroball IV} is used mostly with one or two-fold multiplicities, for
which the total $\gamma$-ray efficiency is very high. The BRS trigger
consists of a kinematical coincidence set-up combining two large-area
heavy-ion telescopes. Both detector telescopes comprise each a
two-dimensional position sensitive low-pressure multiwire chamber in
conjunction with a Bragg-curve ionization chamber. All detection
planes are four-fold subdivided in order to improve the resolution
and to increase the counting rate capability (100 k-events/s). The
two-body Q-value has been reconstructed using events for which both
fragments are in well selected states chosen for spectroscopy
purposes as well as to determine the reaction mechanism responsible
for the population of these peculiar states. The inverse kinematics
of the $^{24}$Mg+$^{12}$C reaction and the negative Q-values give
ideal conditions for the trigger on the BRS, because the chosen
angular range is optimum and because the solid angle transformation
gives a factor~10 for the detection of the heavy fragments. Thus we
have been able to cover a large part of the angular distribution of
the binary process with high efficiency, and a selection of events in
particular angular ranges has been achieved. In binary exit-channels
the exclusive detection of both ejectiles allows precise Q-value
determination, Z-resolution and simultaneously optimal Doppler-shift
correction. \\

\begin{figure}
  \begin{center}
    \includegraphics[width=0.68\textwidth]{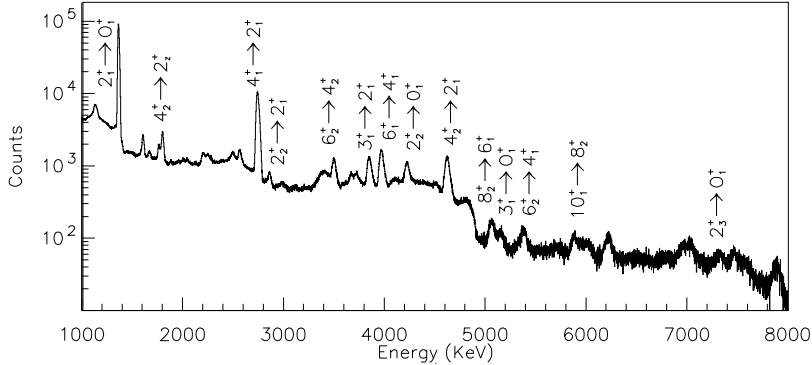}
    \caption{\small\em Doppler corrected $\gamma$-ray spectrum for
                       $^{24}$Mg,
                       using particle-particle-$\gamma$
                       coincidences, measured
                       in the $^{24}$Mg(130 MeV)+$^{12}$C reaction
                       with the BRS/{\sc Euroball IV} detection
system
(see text). }
  \end{center}
  \label{Fig.3}
\end{figure}

Fig.~3 displays a Doppler-corrected $\gamma$-ray spectrum in
coincidence with $^{24}$Mg events identified in the Bragg-Peak vs
energy spectra of the BRS. All known transitions of
$^{24}$Mg~\cite{Beck01,Wiedenhover01} can be identified in the energy
range depicted. As expected we see decays feeding the yrast line of
$^{24}$Mg up to the 8$^{+}_{2}$ level. The population of some of the
observed states, in particular, the 2$^{+}$, 3$^{+}$ and 4$^{+}$
members of the K$^{\pi}$ = 2$^{+}$ rotational band, appears to be
selectively enhanced. The strong population of the K$^{\pi}$ =
2$^{+}$ band and his 4$^{+}$ member at E$_x$ = 6.01 MeV has also been
observed in the $^{12}$C($^{12}$C,$\gamma$) radiative capture
reaction~\cite{Jenkins03}. Furthermore, there is an indication of a
$\gamma$-ray around 5.95 MeV which may be identified with the
10$^{+}_{1}$ $\rightarrow$ 8$^{+}_{2}$ transition as proposed in
Ref.~\cite{Wiedenhover01}. It has been checked in the
$\gamma$-$\gamma$ coincidences that most of the states of Fig.~3
belong to cascades which contain the characteristic 1368 keV
$\gamma$-ray and pass through the lowest 2$^{+}$ state in $^{24}$Mg.
Still a number of transitions in the high-energy part of the spectrum
(6-8~MeV) have not been clearly identified. Even at higher energies,
$^{24}$Mg states appear to show up around 10~MeV (not shown) with
very poor statistics and of unknown structure. Similar states were
also observed in the radiative capture reaction~\cite{Jenkins03}.
Their occurence may be in qualitative agreement with a decay scenario
of radiative capture states proposed by Baye and 
Descouvemont~\cite{Baye84,Descouvemont86} in the framework of a
microscopic study of the $^{12}$C+$^{12}$C system with the Coordinate
Generator Method. The reason why the search for a $\gamma$-decay in
$^{12}$C+$^{12}$C has not been conclusive so
far~\cite{McGrath81,Metag82,Haas97} is due to the excitation energy
in $^{24}$Mg as well as the spin region (8$\hbar$-12$\hbar$) which
were chosen too high. The next step of the analysis will be the use
of the BRS trigger in order to select the excitation energy range by
the two-body Q-value (in the $^{12}$C+$^{24}$Mg channel), and thus we
will be able to study the region around the decay barriers, where
$\gamma$-decay becomes observable. According to recent predictions
$\gamma$-rays from 6$^{+}$ $\rightarrow$ 4$^{+}$ should have
measurable branching ratios. Work is currently in progress to analyse
the $\gamma$ rays from the $^{12}$C($^{24}$Mg,$^{12}$C
$^{12}$C)$^{12}$C ternary breakup reaction.

\section{Summary and conclusions}

We have discussed the possible link between resonant states in
collisions with identical bosons such as $^{12}$C+$^{12}$C,
$^{24}$Mg+$^{24}$Mg and $^{28}$Si+$^{28}$Si and the SD
bands recently discovered in light N=Z nuclei such as $^{36}$Ar,
$^{40}$Ca, $^{48}$Cr, and $^{56}$Ni. A new butterfly mode of
vibration of the well established J$^{\pi}$ = 38$^{+}$ resonance of
the $^{28}$Si-$^{28}$Si triaxial molecule has been discovered
experimentally. The connection of alpha-clustering and quasimolecular
resonances has been discussed with the search for the
$^{12}$C+$^{12}$C molecule populated by the $^{24}$Mg+$^{12}$C
breakup reaction. The most spectacular result is the strong
population of the K$^{\pi}$ = 2$^{+}$ band of the $^{24}$Mg nucleus
that has also been observed in an exploratory investigation of the
$^{12}$C($^{12}$C,$\gamma$) radiative capture 
reaction~\cite{Jenkins03}. Subsequent radiative capture experiments
are planned in the near future with highly efficient spectrometers
(the {\sc Dragon} separator at {\sc Triumf} and the {\sc Fma} at
Argonne) to investigate the overlap of $^{24}$Mg states observed in
the present work with radiative capture states. As far as the
$\gamma$-ray spectroscopy is concerned, the coexistence of
$\alpha$-cluster states and SD states predicted in
$^{32}$S by recent antisymmetrized molecular dynamics (AMD)
calculations~\cite{Horiuchi03} is still an experimental challenge.
This kind of experiments designed to measure very small
$\Gamma$$_\gamma$/$\Gamma$$_{total}$ branching ratios is extremely
difficult since it requires not only high-efficient fragment
detection in conjunction with a high-resolution Ge multidetector such
as the {\sc Gammasphere} and {\sc Euroball} 4$\pi$ $\gamma$ arrays
but also a large amount of beam time. 

\newpage

\noindent
{\small
{\bf Acknowledgments:} I am pleased to acknowledge the physicists
of both the {\sc Icare} and the BRS/{\sc Euroball IV} collaborations,
with special thanks to M. Rousseau, P. Papka, A. S\`anchez i Zafra,
C. Bhattacharya, and S. Thummerer. We thank the staff of the {\sc
Vivitron} for providing us with good stable beams, M.A. Saettel for
preparing targets, and J. Devin and C. Fuchs for their excellent
support during the experiments. Parts of this work was supported by
the french IN2P3/CNRS and the EC Euroviv contract
HPRI-CT-1999-00078.}


\begin{thebibliography}{9}

\bibitem{Svensson00} C.E. Svensson {\it et al.}, Phys. Rev. Lett.
{\bf 85}, 2693 (2000). 

\bibitem{Ideguchi01} E. Ideguchi {\it et al.}, Phys. Rev. Lett. {\bf
87}, 222501 (2001). 

\bibitem{Lenzi96} S.M. Lenzi {\it et al.}, Z. Phys. {\bf A354}, 117
(1996). 

\bibitem{Thummerer01} S. Thummerer {\it et al.}, J. Phys. G: Nucl.
Part. Phys. {\bf 27}, 1405 (2001). 
 
\bibitem{Rudolph99} D. Rudolph {\it et al.}, Phys. Rev. Lett. {\bf
82}, 3763 (1999). 

\bibitem{Inakura02} T. Inakura {\it et al.}, Nucl. Phys. A {\bf 710},
261 (2002). 

\bibitem{Sakuda02} T. Sakuda and S. Ohkubo, Nucl. Phys. A {\bf 712},
59 (2002). 

\bibitem{Kanada02} Y. Kanada-En'yo, K. Kimura, and H. Horiuchi,
AIP Conf. Proc. {\bf 644}, 188 (2003).

\bibitem{Beck04} C. Beck, International Journal of Modern Physics
\bf A\rm, (2004) to be published

\bibitem{Betts81} R.~R. Betts {\it et al.}, \rm Phys. Rev. Lett. \bf
47\rm, 23 (1981).

\bibitem{Beck01} C. Beck {\it et al.}, Phys. Rev. C {\bf 63}, 014607
(2001).

\bibitem{Nouicer99} R. Nouicer {\it et al.}, Phys. Rev. C \bf 60\rm,
041303 (1999). 

\bibitem{Bromley60} K.A. Erb and D.A. Bromley, {\it Treatise on Heavy
Ion Science}, Vol. {\bf 3}, 201 (1985).

\bibitem{Marsh86} S. Marsh and W.D. Rae, Phys. Lett. B {\bf 180}, 185
(1986). 

\bibitem{Flocard84} H. Flocard {\it et al.}, Prog. Theor. Phys. {\bf
72}, 1000 (1984). 

\bibitem{Leander75} G. Leander and S.E. Larsson, Nucl. Phys. A {\bf
239}, 93 (1975). 

\bibitem{Beck94} C. Beck {\it et al.}, Phys. Rev. C \bf 49\rm, 2618
(1994). 

\bibitem{Beck95} C. Beck {\it et al.}, Nucl. Phys. A \bf 583\rm, 269
(1995). 

\bibitem{Chandana02} C. Bhattacharya {\it et al.}, Phys. Rev. C {\bf
65}, 014611 (2002). 

\bibitem{Rousseau02} M. Rousseau {\it et al.}, Phys. Rev. C {\bf 66},
034612 (2002). 

\bibitem{Uegaki94} E. Uegaki and Y. Abe, \rm Phys. Lett. \bf B340\rm,
143 (1994).

\bibitem{Konnerth85} D. Konnerth {\it et al.}, \rm Phys. Rev. Lett.
\bf 55\rm, 588 (1985). 
 
\bibitem{Wuosmaa03} A.~H. Wuosmaa {\it et al.}, \rm Phys. Lett. 
\bf B571\rm, 155 (2003).

\bibitem{Wuosmaa87} A.~H. Wuosmaa {\it et al.}, \rm Phys. Rev. Lett.
\bf 58\rm, 1312 (1987). 

\bibitem{McGrath81} R.L. McGrath {\it et al.}, Phys. Rev. C {\bf 7},
1280 (1981). 

\bibitem{Metag82} V. Metag {\it et al.}, Phys. Rev. C {\bf 25}, 1486
(1982). 

\bibitem{Haas97} F. Haas {\it et al.}, Il Nuovo Cimento {\bf 110A},
989 (1997). 

\bibitem{Beck03} C. Beck {\it et al.}, Nucl. Phys. A (to be
published) and arXiv:nucl-ex/0309007 (2003).

\bibitem{Fulton86} B.R. Fulton {\it et al.}, Phys. Lett. B {\bf
267}, 325 (1991).

\bibitem{Curtis95} N. Curtis {\it et al.}, Phys. Rev. C {\bf 51},
1554 (1995). 

\bibitem{Singer00} S.M. Singer {\it et al.}, Phys. Rev. C {\bf 62},
054609 (2000). 

\bibitem{Wiedenhover01} I. Wiedenh\"over {\it et al.}, Phys. Rev.
Lett. {\bf 87}, 142502 (2001). 

\bibitem{Jenkins03} D.G. Jenkins {\it et al.}, AIP Conf. Proc.
{\bf 656}, 329 (2003); and private communications.

\bibitem{Baye84} D. Baye and P. Descouvemont, Nucl. Phys. A \bf
419\rm, 397 (1984).

\bibitem{Descouvemont86} P. Descouvemont and D. Baye, Phys. Lett.
\bf 196B\rm, 143 (1986).

\bibitem{Horiuchi03} H. Horiuchi, Y. Kanada-En'yo, and K. Kimura,
Nucl. Phys. A {\bf 722}, 80c (2003). 

\end{thebibliography}
\end{document}